\documentclass[superscriptaddress,pra,amsmath,amssymb,eufrak,aps,floatfix,twocolumn,notitlepage,longbibliography]{revtex4-2}

\newcommand{\rpd}[1]{\partial_t #1}

\usepackage{lipsum}
\usepackage{braket}
\usepackage{graphicx}
\usepackage{dcolumn}
\usepackage{enumitem}
\usepackage{bm}
\usepackage{amsmath}
\usepackage{amssymb}
\usepackage{mathtools}
\usepackage{subfigure}
\usepackage{color}
\usepackage{xcolor}
\usepackage{verbatim}
\usepackage{float}
\usepackage{appendix}
\usepackage{bbold}

\newcommand{\abs}[1]{\lvert #1 \rvert} 
\newcommand{\ev}[1]{\langle #1 \rangle} 
\newcommand{\mbf}[1]{\mathbf{#1}} 
\newcommand{\hSig}[0]{\hat{\sigma}} 

\usepackage{upgreek}
\usepackage{physics}
\usepackage{isomath}
\usepackage{placeins}
\usepackage{hhline}
\usepackage{hyperref}
\usepackage{cancel}

\begin{document}

\preprint{APS/123-QED}

\title{Dynamical aspects of steady-state subradiance: Detailed balance and its breakdown}

\author{Athreya Shankar}
\email{athreya@physics.iitm.ac.in}
\affiliation{Department of Physics, Indian Institute of Technology Madras, 600036 Chennai, India}
\affiliation{Center for Quantum Information, Communication and Computing, Indian Institute of Technology Madras, Chennai 600036, India}
\affiliation{Institute for Theoretical Physics, University of Innsbruck, Innsbruck, Austria}
\affiliation{Institute for Quantum Optics and Quantum Information of the Austrian Academy of Sciences, Innsbruck, Austria}
\author{Simon~B.~J\"ager}
\affiliation{Physikalisches Institut, University of Bonn, Nussallee 12, 53115 Bonn, Germany}
\author{Jarrod T. Reilly}
\affiliation{JILA and Department of Physics, University of Colorado, 440 UCB, Boulder, CO 80309, USA}
\author{Raphael Kaubruegger}
\affiliation{JILA and Department of Physics, University of Colorado, 440 UCB, Boulder, CO 80309, USA}
\affiliation{Institute for Theoretical Physics, University of Innsbruck, Innsbruck, Austria}
\affiliation{Institute for Quantum Optics and Quantum Information of the Austrian Academy of Sciences, Innsbruck, Austria}
\author{Murray J. Holland}
\affiliation{JILA and Department of Physics, University of Colorado, 440 UCB, Boulder, CO 80309, USA}
\author{Walter Hahn}
\affiliation{Fraunhofer Institute for Applied Solid State Physics IAF, Tullastr. 72, 79108 Freiburg, Germany}
\affiliation{Institute for Quantum Optics and Quantum Information of the Austrian Academy of Sciences, Innsbruck, Austria}

\date{\today}

\begin{abstract}
The dynamics of dissipative many-body quantum systems sometimes admit an emergent classical description in terms of a Markov chain even though the corresponding state space contains highly entangled states. In particular, a bad-cavity laser is a paradigm system whose dynamics can be formulated as a Markov chain in a two-dimensional state space spanned by collective angular momentum states. In this article, we investigate the connection between a dissipative phase transition that occurs in the subradiant regime of this system in the large atom number limit, and the properties of the underlying Markov chain. In one of the phases, the Markov chain approaches the detailed-balance condition with increasing atom number $N$ and hence becomes effectively time-reversible. This is caused by a collective atomic emission process that effectively reduces the Markov chain to one dimension. In the other phase, we find the emergence of time-asymmetric probability currents in the two-dimensional state space that signals a breakdown of detailed balance. This is accompanied by a macroscopic internal entropy production rate in the Markov chain that scales extensively with the atom number $N$. An observable consequence of the probability currents is a self-pulsing of the cavity light output in this phase, which can be detected as an anti-correlation dip in the intensity correlation function. 
\end{abstract}

\maketitle

\section{Introduction}

Many-body quantum effects that arise from the coupling of atoms to electromagnetic fields have led to many revolutionary theoretical and experimental developments over the past few decades. At the forefront is Dicke's theory of superradiance~\cite{dicke1954PR} which describes how constructive interference causes a group of emitters to coherently radiate light at a rate higher than that of uncorrelated emitters. There also exists a complementary regime in which the atomic emission amplitudes destructively interfere so that the emission rate is greatly suppressed~\cite{dicke1954PR}. This can occur even in the presence of a significant amount of excited state population. In these subradiant states, the weakening of the system's emission rate opens the door to protocols that are at least partially protected from dissipative effects~\cite{ostermann2013PRL}.

Steady-state subradiance can be realized in ``bad-cavity'' or superradiant lasers that are operated below threshold~\cite{meiser2010PRASteadyState,shankar2021PRL,reilly2026PRL}. A minimal model for a bad-cavity laser consists of an ensemble of incoherently pumped atoms that collectively emit into a heavily damped cavity mode~\cite{meiser2009PRL,meiser2010PRASteadyState}. Such subradiant states are remarkable because, in this regime, the atomic system is dominated by noise and dissipation but it nevertheless relaxes to an entangled state that suppresses cavity emission due to destructive interference. While the study of many-body driven-dissipative systems quickly becomes intractable because of the exponential scaling of the underlying Liouville space, here it is possible to map the problem onto a classical Markov chain in a two-dimensional state space that scales only quadratically with the number of atoms~\cite{zhang2018NJP}. The individual states indexed by collective angular momentum quantum numbers $J$ and $M$ correspond to completely mixed states, where the participating states can be highly entangled. Therefore, this laser model serves as an intriguing complex system where tools available to study Markov processes in classical statistical mechanics can be directly applied to a dissipative, interacting many-body quantum system.  

Recent studies of the bad-cavity laser operating below threshold revealed a dissipative phase transition between two distinct phases of steady-state subradiance~\cite{shankar2021PRL}. The phase transition is captured by sharp changes in a number of steady-state observables such as the cavity photon number and the atomic inversion as the number of atoms $N$ coupling to a fixed cavity mode is increased. In fact, these observables exhibit non-analytic changes at the critical point in the limit  $N\to \infty$. The many-body steady state near the critical region was shown to be a highly entangled state, with the fraction of unentangled atoms becoming vanishingly small as $N\to\infty$.    

In this article, we unravel \emph{dynamical} aspects of this subradiant-to-subradiant phase transition by focusing on the dynamics of the underlying Markov chain. We show that, at the critical point, the Markov chain changes from a time-reversible chain to an irreversible one as the pump rate is increased. This change is evidenced by a breakdown of detailed balance, which we characterize using the internal entropy production rate~\cite{Gaspard,Seifert_2012,esposito,seifert2005PRL,cocconi2020Entropy,landi2021RMP}. At the critical point, we observe the onset of an extensive entropy production rate indicating the emergence of time-irreversible probability currents in the two-dimensional state space. These currents, in turn, result in an observable self-pulsing of the light output from the cavity~\cite{temnov2009OptExp}. Moreover, we show that the pulsing is captured in a dynamical observable, namely, the intensity correlation function $g^2(\tau)$, which is bunched at zero time delay ($\tau=0$) but shows sub-Poissonian anticorrelation dips at certain delays $\tau>0$. In the time-reversible phase, detailed balance is preserved by an effective reduction of the state space to one dimension by the dominant collective-emission process. 

We remark that the development of a unified theory of entropy production that applies to both classical and quantum physics is an active area of research~\cite{landi2021RMP}. For quantum systems, phase-space approaches to study entropy production have been proposed~\cite{santos2017PRL,santos2018PRA}. Such methods have been used to study driven-dissipative systems such as the open Kerr  and Dicke models~\cite{goes2020PRR}. However, in this work, the emergent description of the quantum master equation as a Markov chain allows us to use the framework of classical entropy production to characterize this quantum many-body system.

The layout of this paper is as follows. In Sec.~\ref{sec:model}, we introduce our model for a bad-cavity laser and a rate equation formalism for obtaining exact solutions. We also provide physical insight into the state of the constituent atoms in the collective $J,M$ states underlying the Markov chain. In Sec.~\ref{sec:ss_phases}, we explore the steady-state population distribution in these phases and qualitatively discuss the time reversibility of the Markov chain. We quantify this notion using the entropy production rate in Sec.~\ref{sec:epr}. In  Sec.~\ref{sec:intensity},  we show how the phase transition manifests as observable changes in the second-order coherence function  and conclude with a summary in Sec.~\ref{sec:conc}.

\section{\label{sec:model} Model and methods}

\begin{figure}[!tb]
    \centering
    \includegraphics[width=0.75\columnwidth]{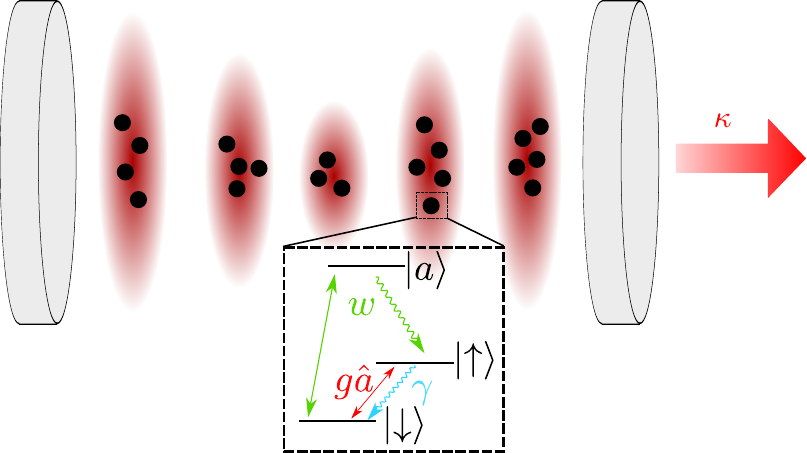}
    \caption{Schematic of a bad-cavity laser with atoms trapped at the antinodes of a lossy cavity mode. The level structure of each atom is shown in the inset.}
    \label{fig:setup}
\end{figure}

We consider $N$ atomic dipoles with free-space spontaneous emission rate $\gamma$ to be trapped at the antinodes of a cavity mode as shown in Fig.~\ref{fig:setup}. We assume that the atomic transition $\ket{\uparrow}\leftrightarrow\ket{\downarrow}$ is resonant with the cavity mode and exchanges excitations with it via the Jaynes-Cummings interaction, $\hat{H}_\text{JC}=\sum_{j=1}^N (g/2)\left(\hat{a}^\dag \hSig_j^- + \hat{a}\hSig_j^+\right)$. Here $\hat{a}^\dag,\hat{a}$ are the creation and annihilation operators for the mode. The $j$-th two-level system formed by $\ket{\uparrow}_j,\ket{\downarrow}_j$ is described by Pauli matrices defined by $\hSig_j^-=\ket{\downarrow}_j\bra{\uparrow}_j = (\hSig_j^+)^\dag$ and $\hSig_j^z=\ket{\uparrow}_j\bra{\uparrow}_j-\ket{\downarrow}_j\bra{\downarrow}_j$. When the cavity loss rate $\kappa$ is large compared to the collective atom-cavity coupling strength $\sqrt{N} g$ and~$\gamma$, the system is in a bad-cavity limit and the cavity mode can be eliminated~\cite{xu2016theory,jaeger2022PRL}. The effective dynamics of the atoms then corresponds to collective emission through the cavity with Lindblad jump operator $\sqrt{\Gamma_c}\hat{J}^-$. Here, we have introduced $\Gamma_c = g^2/\kappa$ and the collective angular momentum operators $\hat{J}^\pm=\sum_j \hSig_j^\pm$. While atoms emit indistinguishably into the cavity mode, we assume the emission into free space to be distinguishable, since the inter-atom distance is larger than the wavelength of the emission making it in principle possible to uniquely identify the emitting atom. Hence, this process is modeled by jump operators $\sqrt{\gamma}\hSig_j^-$, one for each atom. The atoms are incoherently and non-collectively repumped ($\ket{\downarrow}\rightarrow\ket{\uparrow}$) at effective rate $w$, which is described by the jump operators $\sqrt{w}\hSig_j^+$. A simple example of a repumping scheme is shown in Fig.~\ref{fig:setup}, where a coherent drive couples $\ket{\downarrow}$ to an auxiliary state $\ket{a}$ which then rapidly decays to $\ket{\uparrow}$. The master equation for the effective atomic dynamics after eliminating the cavity mode is given by~\cite{PhysRevLett114}
\begin{eqnarray}
    \rpd \hat{\rho} &=& \sum_{j=1}^N \left( \hat{\mathcal{D}}\left[\sqrt{w} \hSig_j^{+}\right] \hat{\rho} + \hat{\mathcal{D}}\left[\sqrt{\gamma} \hSig_j^{-}\right] \hat{\rho} \right) + \hat{\mathcal{D}}\left[\sqrt{\Gamma_c} \hat{J}^-\right] \hat{\rho} \nonumber\\
    &\equiv& \mathcal{L}\hat{\rho},
    \label{eqn:me}
\end{eqnarray}
where $\hat{\mathcal{D}}[\hat{O}] \hat{\rho} = \hat{O} \hat{\rho} \hat{O}^{\dagger} -  \hat{O}^{\dagger} \hat{O}\hat{\rho}/2 - \hat{\rho} \hat{O}^{\dagger} \hat{O}/2$ is the Lindblad dissipator associated with a jump operator $\hat{O}$.

The master equation~(\ref{eqn:me}) is invariant under the permutation of particles. This permutation symmetry greatly reduces the number of independent basis states in Liouville space required to describe the evolution of an initially permutation invariant density matrix from $4^N$ to $\mathcal{O}(N^3)$~\cite{PhysRevA87,zhang2018NJP,PhysRevA98}. To write down these basis states, we first note that the angular momentum of $N$ spin-1/2 particles can be combined in multiple ways to obtain pure states with a fixed value of total angular momentum $J$ and $z$-component $M$. The multiplicity of states for a fixed $J$ and $M$ is given by $d_N^J = N! (2J + 1)/[(N/2 + J + 1)! (N/2-J)!]$~\cite{mandel1995Optical}. Permutation invariant basis states in Liouville space $\hat{S}_{J,M,M'}$ are described by the equal mixture of these pure states~\cite{chase2008PRA}
\begin{eqnarray}
\hat{S}_{J,M,M'} = \frac{1}{d_N^J}\sum_{\chi=1}^{d_N^J} \ket{J,M,\chi}\bra{J,M',\chi},
\label{eqn:s_jmmp}
\end{eqnarray}
where the auxiliary quantum number $\chi=1,2,\ldots,d_N^J$ distinguishes the degenerate basis states with the same $J,M$ values. The total angular momentum can take values $0,1,\ldots,N/2$ for an even number $N$, while the indices $M,M'$ take values in the range $-J,-J+1,\ldots,J$ for a given $J$. The result is a total of $\mathcal{O}(N^3)$ basis states. Furthermore, the master equation~(\ref{eqn:me}) is invariant under a $U(1)$ transformation $\hSig_j^\pm \rightarrow e^{\pm i\phi}\hSig_j^\pm$, which implies that the dynamics allows transitions only between states with the same value of $M-M'$. This results in an additional reduction in the number of basis states to $(N+2)^2/4\sim \mathcal{O}(N^2)$ and makes the $M'$ index redundant. Introducing $\hat{S}_{J,M}\equiv\hat{S}_{J,M,M}$, the particle permutation invariant and $U(1)$ symmetric density matrices can therefore be described as 
\begin{eqnarray}
\hat{\rho}_s = \sum_{J,M}P_{J,M}\hat{S}_{J,M},
\end{eqnarray}
with probability conservation $\sum_{J,M} P_{J,M}=1$.

The master equation in Eq.~\eqref{eqn:me} can be recast as a rate equation for the evolution of the probabilities $P_{J,M}$:
\begin{eqnarray}
\frac{d}{dt}P_{J,M} = \sum_{(J',M') \neq (J,M)} &&\left( R_{J,M;J',M'}P_{J',M'} \right. \nonumber\\
&&\left. -R_{J',M';J,M}P_{J,M}\right),
\label{eqn:rate_eqn}
\end{eqnarray}
where $R_{J,M;J',M'}$ is the total transition rate from $J',M'\rightarrow J,M$. The expressions for these rates were derived in Ref.~\cite{chase2008PRA,baragiola2010PRA} and are summarized in Ref.~\cite{zhang2018NJP}. For convenience, we reproduce these expressions in Appendix~\ref{app_a}. We thus arrive at an effective description of the quantum dynamics governed by Eq.~\eqref{eqn:me} as a classical Markov chain. We find the steady-state population distribution as the solution to the system of linear equations given by  $\frac{d}{dt}P_{J,M}=0 \;\forall\; J,M$. 
The properties of the Markov chain~(\ref{eqn:rate_eqn}) are discussed in Sections~\ref{sec:ss_phases} and~\ref{sec:epr}  and constitute the main result of this paper. Before proceeding, however, we first discuss the nature of the states underlying this Markov chain.

\subsection{\label{sec:pi_basis}Microscopic nature of permutation invariant basis states}

Although we only require basis states $\hat{S}_{J,M,M'}$ with $M=M'$ for our work, in the following we relax this restriction for a more general discussion. The collective angular momentum formulation, while very useful, does not provide intuition about the state of the individual constituent atoms.  Previous works~\cite{chase2008PRA} have rigorously established the connection between the permutation invariant (PI) collective angular momentum states and the individual atomic states, but extracting an intuitive understanding from the formal treatment is not straightforward. Here, we bridge this gap by motivating the explicit form of the permutation invariant basis states in terms of the constituent atoms.

Let us first consider the $J=N/2$ manifold. The pure states in this manifold are the well known Dicke states, given by the permutation symmetric \emph{superposition}
\begin{eqnarray}
    \ket{N,M} = \left(\ket{\uparrow}^{\otimes N/2+M}\ket{\downarrow}^{\otimes N/2-M}\right)_\text{PI}.
\end{eqnarray}
The subscript PI indicates that all permutations of the atomic indices are summed over and the resulting state is normalized. The corresponding Liouville basis states $\hat{S}_{N/2,M,M'}$ are simply given by $\hat{S}_{N/2,M,M'}=\ket{N,M}\bra{N,M'}$ and represent populations ($M=M'$) in or coherences ($M\neq M'$) between the various Dicke states. 

Next, let us consider the PI $J=0$ state $\hat{S}_{0,0,0}$. In Ref.~\cite{urizar2013PRA}, this state was shown to be a uniform mixture of pairwise singlets. To provide an explicit expression for this state, let us define $\ket{\psi_{a,b}}_s$ as the singlet state composed of spins $a,b$, i.e.
$\ket{\psi_{a,b}}_s = \left(\ket{\uparrow}_a\ket{\downarrow}_b - \ket{\downarrow}_a\ket{\uparrow}_b \right)/\sqrt{2}$.
The corresponding density matrix is $\hat{\rho}_{a,b}^s = \ket{\psi_{a,b}}_s \bra{\psi_{a,b}}_s$. Then, the PI $J=0$ state can be expressed as 
\begin{eqnarray}
	\hat{S}_{0,0,0} = \left(\hat{\rho}_{12}^s \otimes \hat{\rho}_{34}^s \otimes \ldots \otimes \hat{\rho}_{N-1,N}^s\right)_\text{PI},
\end{eqnarray}  
representing the fact that this state consists of a mixture of $N/2$ singlet pairs. 

We now combine the above observations to write down the form of the states for $0<J<N/2$. We propose that any general $\hat{S}_{J,M,M'}$ state has the form 
\begin{eqnarray}
	\hat{S}_{J,M,M'} = &&\left(\hat{\rho}_{12}^s\otimes\rho_{34}^s\otimes \ldots\otimes \hat{\rho}_{N-2J-1,N-2J}^s \otimes \right.\nonumber\\
	&&\left.\left\lvert 2J,M \right\rangle \left \langle 2J,M'\right\rvert  \right)_\text{PI},
	\label{eqn:form}
\end{eqnarray} 
i.e. out of the $N$ atoms, $N-2J$ atoms are in singlet pairs, while the remaining $2J$ atoms are in their maximal angular momentum manifold (indexed by $J$). As before, the $M=M'$ states represent populations in the Dicke states in this manifold, while the $M\neq M'$ states represent coherences between them.  In Appendix~\ref{app:proof_pi}, we straightforwardly show that the RHS of Eq.~(\ref{eqn:form}) satisfies the following equations:
\begin{eqnarray}
	\hat{\mbf{J}}^2 \hat{S}_{J,M,M'} &=& J(J+1)\hat{S}_{J,M,M'}, \nonumber\\ \hat{J}_z\hat{S}_{J,M,M'} &=& M \hat{S}_{J,M,M'}, \nonumber\\
	\hat{S}_{J,M,M'} \hat{J}_z &=& M'\hat{S}_{J,M,M'}.
	\label{eqn:props}
\end{eqnarray} 
In addition, as the proposed form is explicitly PI, it correctly describes the states of the constituent atoms in each of the PI basis states $\hat{S}_{J,M,M'}$.

\section{\label{sec:ss_phases} Subradiant phases}

\begin{figure}[!tb]
    \centering
    \includegraphics[width=0.7\columnwidth]{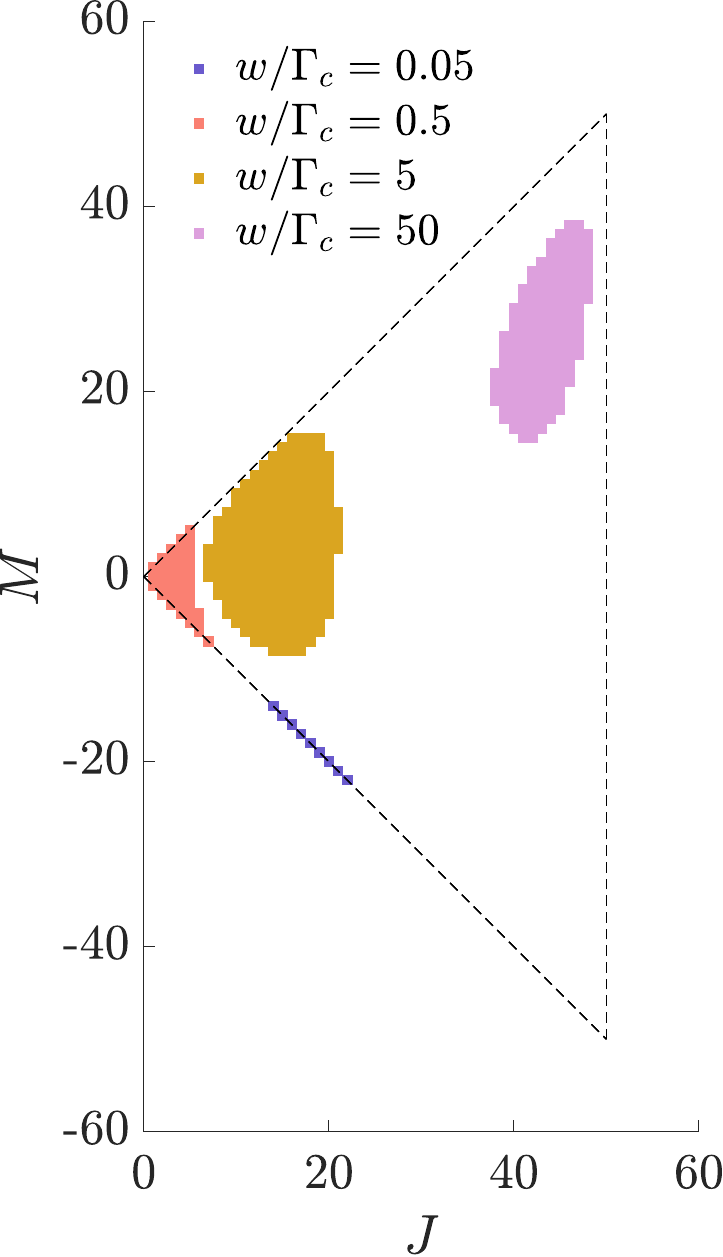}
    \caption{Steady-state population distribution for $N=100$ atoms and for different values of $w$. The colored areas depict the highest occupied states that contribute to $\sim 68\%$ of the steady-state population.}
    \label{fig:n100_multimapn}
\end{figure}

We now explore the steady-state and dynamical properties of the subradiant phases of a bad-cavity laser. For fixed emission rates into free space $\gamma$ and into the cavity $\Gamma_c$, a dissipative phase transition between two subradiant phases occurs at the critical repump rate $w=\gamma$~\cite{shankar2021PRL}. We fix $\gamma/\Gamma_c=0.1$ in this work. Let us first consider the case $w<\gamma$. As illustrated in Fig.~\ref{fig:n100_multimapn}, the steady-state population is essentially concentrated on the lower boundary of the triangular state space where $M=-J$ because the collective decay dominates the individual atom processes and suppresses dynamics taking the system away from this boundary. The steady-state distribution along the boundary is determined by the balance of upwards hopping driven by the repump and downward hopping driven by spontaneous emission. This hopping occurs only between nearest neighbor states.

From the results of the previous section, we can understand the steady state in this regime as a mixture of atoms in $\ket{\downarrow}$ and in singlet pairs. Both of these states cannot emit into the
cavity, and therefore the system in this phase is
almost perfectly dark with respect to emission into the
cavity. A small non-zero emission nevertheless occurs because of rare events wherein a single collective atomic excitation is induced and the system leaves the lower boundary. 

Denoting the population in a state $(J,M)=(J,-J)$ by $P_J$, we find (see Appendix~\ref{app:pop_an}) that the steady-state population satisfies 
\begin{equation}
	P_J = \frac{R_{J-1}^+}{R_J^-}P_{J-1},
	\label{eqn:pj_recursion_main}
\end{equation}
where the rates $R_J^\pm$ are detailed in Appendix~\ref{app:pop_an}. Assuming a continuum limit for index $J$, we find that the steady-state distribution is a Gaussian distribution with mean $\mu$ and variance $\sigma^2$ given by 
\begin{eqnarray}
\mu = \frac{N}{2}\left[\frac{\gamma-w}{\gamma+w}\right],\;
\sigma^2 = \frac{N\gamma w}{(\gamma+w)^2}.
\end{eqnarray}

As $w$ approaches $\gamma$ from below ($w\rightarrow \gamma^-$), the population is very close to the vertex $J=0,M=0$, but still on the lower boundary, with a spread that is only $\mathcal{O}(\sqrt{N})$. Since the $J=0$, $M=0$ state describes a mixture of singlet pairs, this implies that an increasing number of atoms begin to form singlet pairs as $w\rightarrow\gamma^-$. 

As an aside, in Appendix~\ref{app:pop_an}, we also analytically determine the steady-state distribution in the case of $\gamma=0$ and $w\rightarrow 0$. Interestingly, in this limit, the steady-state population is independent of $N$ and $w$ (for $N\gg 1$). The population is again concentrated in states with $M=-J$ but only $J=0,1,2$ are significantly populated with roughly $21\%$, $62\%$ and $17\%$ probability of occupation respectively.

The recurrence relation Eq.~(\ref{eqn:pj_recursion_main}) is the condition for detailed balance in a Markov chain. This implies that the Markov chain in the regime $w<\gamma$ is time reversible. We note that, in general, a steady-state unidirectional population current cannot be sustained in a finite 1D Markov chain with only nearest neighbor hopping, and therefore all such chains are necessarily time reversible.

We now consider the phase when $w>\gamma$. As shown in Fig.~\ref{fig:n100_multimapn}, the steady-state population for $\gamma<w<\gamma+\Gamma_c$  is localized in the region close to the apex of the triangle, where $J\sim 0$ and $M\sim 0$. The point $w=\gamma+\Gamma_c$ marks the onset of steady-state superradiance; further increasing $w$ beyond this point leads to the steady-state population being concentrated in the interior of the triangular state space. For $w>\gamma$, collective states with $M>-J$ are significantly occupied. Hence, as discussed in the previous section, a fraction of the atoms are in Dicke states that collectively emit into the cavity, giving rise to light output from the cavity that scales extensively with the atom number $N$~\cite{shankar2021PRL}.  

\begin{figure}[!tb]
    \centering
    \includegraphics[width=0.85\columnwidth]{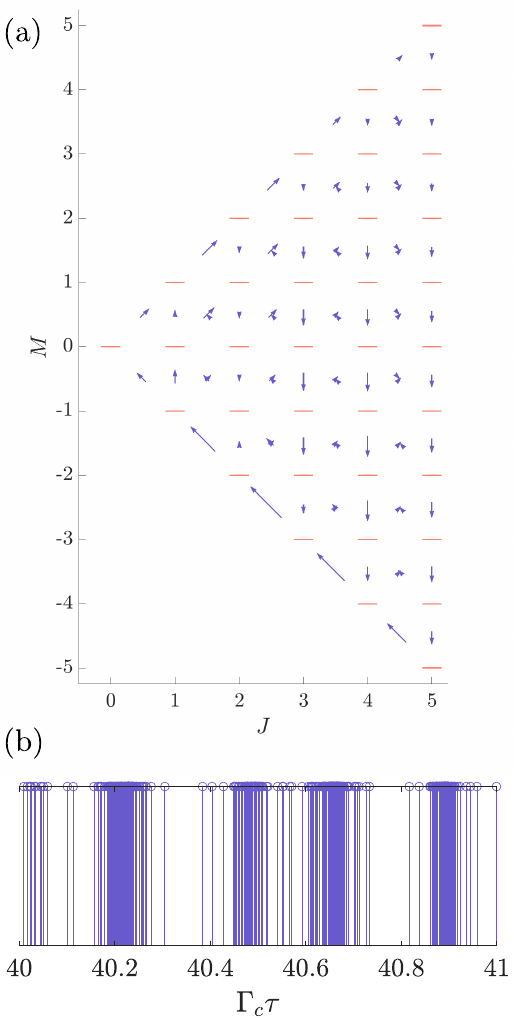}
    \caption{(a) Population flow in steady state for the case of $N=100$ and $w/\Gamma_c=0.2$ showing the existence of current loops associated with breaking of time-reversal symmetry. (b) Record of collective emission jump times in a simulated trajectory of the Markov chain showing intermittent bursts followed by periods of inactivity. Here, $N=10^4$ and $w/\Gamma_c=0.2$. The initial time $\Gamma_c\tau=40$ in this plot is chosen to ensure that the system has relaxed to the steady state, which requires timescales $\Gamma_c\tau\gg 1$. }
    \label{fig:arrow_photodet}
\end{figure}

Importantly, the population for $w>\gamma$ is not confined to a line, and hence the 2D state space cannot be reduced to a 1D chain as in the $w<\gamma$ phase. In contrast to the latter phase, we find a breakdown of detailed balance in this regime by investigating the net probability currents
\begin{eqnarray}
W_{J,M;J',M'}=P_{J,M}R_{J^\prime,M^\prime;J,M}-P_{J^\prime,M^\prime}R_{J,M;J^\prime,M^\prime}
\label{eqn:nps}
\end{eqnarray}
between every pair of nearest neighbors. Numerical results for the net probability currents shown in Fig.~\ref{fig:arrow_photodet}(a) clearly demonstrate the presence of probability-current loops and gives insight into the temporal dynamics of the system in this phase: As repump dominates over spontaneous emission, population initially concentrated on the lower boundary climbs up this boundary, proceeds further along the upper boundary until sufficient inversion is built up for a cascade to occur via collective emission. The cycle then repeats as the cascade returns the population to the lower boundary. In Fig.~\ref{fig:arrow_photodet}(b), we show the jump times associated with the collective emission process in a simulated trajectory of the Markov chain at long times after it has reached steady state. Although such trajectories do not correspond to monitored dynamics under real physical measurements~\cite{zhang2018NJP}, the simulation suggests that the cavity output is intermittent with alternating periods of bursting and no output. Indeed, in Sec.~\ref{sec:intensity}, we will confirm this self-pulsing behavior of the cavity output for $w>\gamma$ by computing the intensity correlation function.

\section{\label{sec:epr}Entropy production rate}

In this section, we quantify the breakdown of detailed balance, and hence the extent of time-irreversibility in the underlying Markov chain, using the internal entropy production rate $\dot{S}_i$~\cite{Gaspard,Seifert_2012,esposito,cocconi2020Entropy}. We briefly recall its definition, closely following Ref.~\cite{cocconi2020Entropy}. At any time $t$, the Shannon entropy of the population distribution in the  Markov chain is given by 
\begin{eqnarray}
S_\text{tot}(t) = -\sum_{J,M} P_{J,M}(t) \ln \left[ P_{J,M}(t) \right].
\end{eqnarray}
Taking the time derivative and using the rate equation~(\ref{eqn:rate_eqn}), the total rate of entropy production can be written as  
\begin{eqnarray}
\dot{S}_\text{tot} = &&\frac{1}{2}\sum_{(J',M') \neq (J,M)} W_{J,M;J',M'} \ln \left( \frac{P_{J,M}}{P_{J',M'}} \right),
\label{eqn:s_tot}
\end{eqnarray}
where $W_{J,M;J',M'}$ is given by Eq.~(\ref{eqn:nps}). We have dropped the time dependency of $P_{JM}$ in Eq.~(\ref{eqn:s_tot}) because we consider the entropy production rate in steady state. We can separate $\dot{S}_\mathrm{tot}$ into  $\dot{S}_\mathrm{tot} = \dot{S}_e + \dot{S}_i$~\cite{cocconi2020Entropy,seifert2005PRL}, where the external part $\dot{S}_e$ and internal part $\dot{S}_i$ are respectively given by 
\begin{eqnarray}
\dot{S}_e&\!= &\!\frac{1}{2}\sum_{(J',M') \neq (J,M)} \!\!W_{J,M;J',M'} \ln\!\left( \frac{R_{J,M;J',M'}}{R_{J',M';J,M}} \right)\!,\nonumber\\ 
\dot{S}_i&\! = &\!\frac{1}{2}\sum_{(J',M') \neq (J,M)}\!\!W_{J,M;J',M'} \ln\!\left(\frac{P_{J,M}R_{J^\prime,M^\prime;J,M}}{P_{J^\prime,M^\prime}R_{J,M;J^\prime,M^\prime}} \right)\!. \nonumber\\
\end{eqnarray}
A nonvanishing $\dot{S}_i$ in steady-state is a signature of irreversibility in the Markov chain~\cite{Gaspard,Seifert_2012} and the external part $\dot{S}_e$ is usually referred to as the entropy flow to the environment~\cite{esposito}.

For any time $t>0$, while $\dot{S}_e(t)$ can be negative or positive, $\dot{S}_i(t) \geq 0$ because each term in its summation is of the form $(x-y)\ln(x/y) \geq 0$ for $x,y>0$~\cite{cocconi2020Entropy}. Furthermore, in steady state, both $\dot{S}_e$ and $\dot{S}_i$ are zero for a Markov chain that satisfies detailed balance, whereas only the combination $\dot{S}_e + \dot{S}_i$ vanishes for a chain with broken detailed balance. This can be understood by noting that each term in the summation for $\dot{S}_e$ and $\dot{S}_i$ is proportional to the net probability current $W_{J,M;J',M'}$, which vanishes for each pair of states under the condition of detailed balance [Eq.~(\ref{eqn:nps})]. The internal entropy production rate can also be viewed as a Kullback-Leibler divergence of the probability distributions for trajectories in the $J,M$ state space and the corresponding time-reversed trajectories~\cite{Gaspard,cocconi2020Entropy}. Therefore, it is a measure of whether the arrow of time can be determined based on individual trajectories~\cite{edgar2015PRL}. 

\begin{figure}[!tb]
    \centering
    \includegraphics[width=0.9\columnwidth]{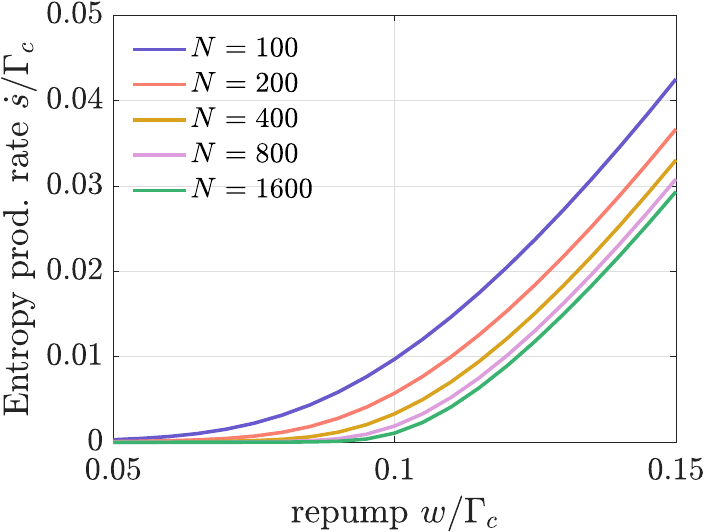}
    \caption{Internal entropy production rate (normalized to system size) as a function of repump for different system sizes.  Here, $w=\gamma=0.1 \Gamma_c$ is the critical point indicated by the vertical line.}
    \label{fig:entropy_prod_rate}
\end{figure}

In Fig.~\ref{fig:entropy_prod_rate}, we plot the internal entropy production rate per atom $\dot{s}=\dot{S}_i/N$ as a function of the repump rate $w$ for different atom numbers $N$. With increasing $N$, we find evidence for the emergence of a sharp change in $\dot{s}$ as $w$ crosses $\gamma$. Specifically, as $N$ increases, $\dot{s}\approx 0$ for $w<\gamma$ while it saturates to a non-zero constant for any fixed $w>\gamma$, indicating macroscopic entropy production in the latter phase. These observations demonstrate that the subradiant-to-subradiant phase transition is accompanied by the breaking of time-reversal symmetry. This also illustrates the power of the rate equation approach, which allows us to analyze a quantum many-body system using tools developed for classical Markov chains and to study the dissipative dynamics of quantum states of over 1000 atoms. 

\section{\label{sec:intensity}Intensity fluctuations}

As discussed in Sec.~\ref{sec:ss_phases}, the absence of time-reversal symmetry for $w>\gamma$ translates into observable fluctuations of the intensity of the output light. In this section, we explore how these intensity fluctuations manifest in the two-photon intensity correlation function defined as 
\begin{eqnarray}
g^2(\tau) = \frac{\ev{\hat{J}^+(t)\hat{J}^+(t+\tau)\hat{J}^-(t+\tau)\hat{J}^-(t)}}{\ev{\hat{J}^+(t)\hat{J}^-(t)}^2},
\end{eqnarray}
where $\ev{\ldots}$ denotes the quantum mechanical expectation value and $t\gg 1/w$ is a sufficiently long time after which a steady state is established. To compute $g^2(\tau)$ for $\tau>0$, we note that the numerator can be expressed as 
\begin{eqnarray}
\ev{\hat{J}^+(t)\hat{J}^+(t+\tau)\hat{J}^-(t+\tau)\hat{J}^-(t)} = \nonumber\\ \mathrm{Tr}\left[\hat{J}^+\hat{J}^- e^{\mathcal{L}\tau} (\hat{J}^- \hat{\rho}_\text{ss} \hat{J}^+)\right],
\end{eqnarray}
where $\mathcal{L}$ is the Liouvillian governing the system (see Eq.~(\ref{eqn:me})) and $\hat{\rho}_\text{ss}$ is the steady-state density matrix. Therefore, the numerator can be computed as the expectation value of $\hat{J}^+\hat{J}^-$ with respect to a time-evolved auxiliary matrix which is initially given by $\hat{J}^- \hat{\rho}_\text{ss} \hat{J}^+$.

\begin{figure}[b]
    \centering
    \includegraphics[width=0.75\columnwidth]{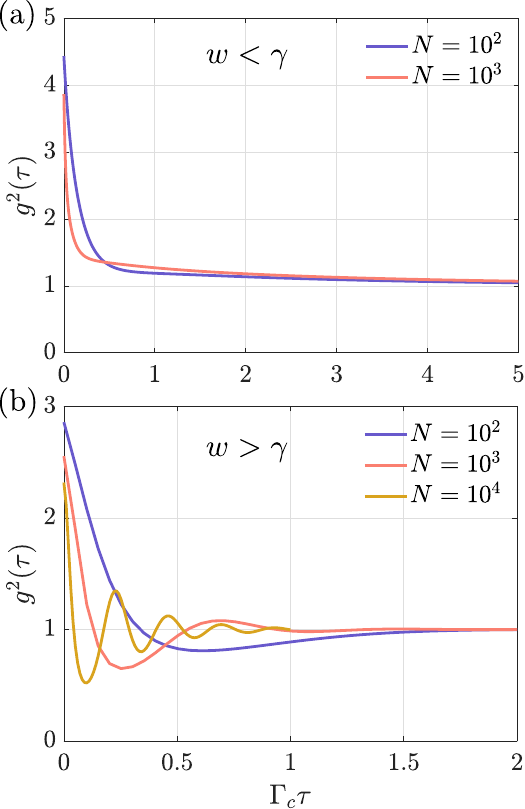}
    \caption{Two-photon intensity correlation function $g^2(\tau)$ at time delay $\tau$ for different system sizes, and for (a) $w/\Gamma_c=0.09$ and (b) $w/\Gamma_c=0.2$.}
    \label{fig:g2Tau}
\end{figure}

Figure~\ref{fig:g2Tau} shows $g^2(\tau)$ as a function of $\tau$ in the two subradiant phases for different system sizes. For $\tau=0$, both phases display strong photon bunching that is characteristic of the subradiant regime. However, for non-zero time delay, the $g^2(\tau)$ function is monotonically decreasing for $w<\gamma$ while it shows oscillatory behavior for $w>\gamma$ with sub-Poisonnian intensity fluctuations ($g^2(\tau)<1$) in certain intervals of $\tau$. The peaks and dips in the correlation function in the latter case can be understood in the light of Fig.~\ref{fig:arrow_photodet} as a self-pulsing cavity output with cycles of alternating bursts and dark periods. Hence, measurement of the $g^2(\tau)$ function can be used as an indicator of the phase transition between the time-reversible and time-irreversible phases. Finally, we note that the self-pulsing observed here is not a mean-field effect and arises as a result of quantum noise~\cite{temnov2009OptExp}. 

\section{\label{sec:conc} Conclusion}
By using an emergent classical description for the bad-cavity laser, we uncover dynamical aspects of the subradiant-to-subradiant phase transition. The classical description is given by a Markov chain, which lives in a triangular state space spanned by the collective angular momentum quantum numbers $J,M$. We show that, above the critical point, the dynamics in the Markov chain becomes time irreversible. The time irreversibility is due to the emergence of population currents in the steady-state of the Markov chain and leads to a non-vanishing entropy production rate per atom in the steady state. These population currents have observable consequences in the form of self-pulsed light output and sub-Poissonian intensity fluctuations that are captured by the $g^2(\tau)$ function. Below the critical point, the dynamics is confined to the lower boundary of the triangle and the effective 1D Markov chain satisfies detailed balance, resulting in a vanishing entropy production rate per atom with increasing system size.

Our work illustrates the possibility of using classical tools of nonequilibrium statistical mechanics to study dissipative phase transitions in certain dissipative many-body quantum systems whose dynamics can be formulated as rate equations. In the future, it will be interesting to leverage the rate-equation formulation to study the transient dynamics of the system, when the repump rate is quenched across the critical point, and thereby explore the sudden switching on or off of the cavity light output as the system changes phase.

\section*{Data and Code Availability}

The code used to generate the figures is publicly available~\cite{dataset}.

\begin{acknowledgments}
We thank Peter Zoller for helpful discussions and feedback on the manuscript. A.S. acknowledges support by the Department of Science and Technology, Govt. of India through the INSPIRE Faculty Award (DST/INSPIRE/04/2023/001486), by the Anusandhan National Research Foundation (ANRF), Govt. of India through the Prime Minister’s Early Career Research Grant (PMECRG) (ANRF/ECRG/2024/001160/PMS) and by IIT Madras through the New Faculty Initiation Grant (NFIG). S.B.J acknowledges support from the Deutsche Forschungsgemeinschaft (DFG, German Research Foundation) under 6 Project No. 277625399-TRR 185 OSCAR (“Open System Control of Atomic and Photonic Matter”, B4) and under Germany’s Excellence Strategy – Cluster of Excellence Matter and Light for Quantum Computing (ML4Q) EXC 2004/1 – 390534769. J. T. R. and M. J. H. acknowledge support from NSF Grant No. 2016244 and NSF Grant No. 2317149. W. H. acknowledges support
from a research fellowship from the DFG (Grant No. HA 8894/1-1).
\end{acknowledgments}

\appendix

\section{Table of rate matrix elements}\label{app_a}

Here, we provide the rate matrix elements that enter Eq.~(\ref{eqn:rate_eqn}) in Table~\ref{tab:rates} (see also Refs.~\cite{shankar2021PRL,zhang2018NJP}).

\renewcommand{\arraystretch}{2.5}
\setlength{\tabcolsep}{12pt}

\begin{table*}[!tb]
    \centering
    \begin{tabular}{ccc}
    Process &   Transition  &   Rate \\\hline
    Collective decay     &  $(J,M)\rightarrow(J,M-1)$   & $\Gamma_c(J+M)(J-M+1)$  \\
    Individual repump     &   $(J,M)\rightarrow(J,M+1)$   &   $w\left[\dfrac{(N+2)(J-M)(J+M+1)}{4J(J+1)}\right]$   \\
        &   $(J,M)\rightarrow(J-1,M+1)$   &   $w\left[\dfrac{(N+2J+2)(J-M)(J-M-1)}{4J(2J+1)}\right]$   \\
        &   $(J,M)\rightarrow(J+1,M+1)$   &   $w\left[\dfrac{(N-2J)(J+M+1)(J+M+2)}{4(J+1)(2J+1)}\right]$   \\ 
    Individual decay     &   $(J,M)\rightarrow(J,M-1)$   &   $\gamma\left[\dfrac{(N+2)(J+M)(J-M+1)}{4J(J+1)}\right]$   \\
        &   $(J,M)\rightarrow(J-1,M-1)$   &   $\gamma\left[\dfrac{(N+2J+2)(J+M)(J+M-1)}{4J(2J+1)}\right]$   \\
        &   $(J,M)\rightarrow(J+1,M-1)$   &   $\gamma\left[\dfrac{(N-2J)(J-M+1)(J-M+2)}{4(J+1)(2J+1)}\right]$   \\
    \end{tabular}
    \caption{Transition rates for different Lindblad terms entering the master equation for the bad-cavity laser~\cite{zhang2018NJP}.}
    \label{tab:rates}
\end{table*}

\section{Proof of the proposed form of PI basis states}
\label{app:proof_pi}
Here, we explicitly demonstrate that the microscopic form of the PI basis states proposed in Eq.~(\ref{eqn:form}) satisfies the conditions presented in Eq.~(\ref{eqn:props}). 

Let us consider a particular component of the basis $S_{J,M,M'}$:
\begin{eqnarray}
	\hat{T}_1 = &&\left(\hat{\rho}_{12}^s\otimes\hat{\rho}_{34}^s\otimes \ldots\otimes \hat{\rho}_{N_J-1,N_J}^s\right)_\text{PI} \nonumber\\
	&&\otimes  \left\lvert 2J,J,M \right\rangle \left \langle 2J,J,M'\right\rvert.
\end{eqnarray}
We label the set of indices in the singlet pairs as \mbox{$A=\left\{1,2,\ldots,N_J\right\}$}, and those in the Dicke state as \mbox{$B=\left\{N_J+1,\ldots,N\right\}$}. Other components are obtained by having a different set of $N_J$ indices in $A$ and $2J$ indices in $B$. We can write 
\begin{eqnarray}
\hat{\mbf{J}} = \hat{\mbf{J}}^{(A)} + \hat{\mbf{J}}^{(B)}.  
\end{eqnarray}
The action of $\hat{\mbf{J}}^{(A)}$ on $\hat{T}_1$ is zero, because the $N_J$ atoms in $A$ are in the $J_A=0$ PI basis state. On the other hand, we can immediately see that 
\begin{eqnarray}
	\left(\hat{\mbf{J}}^{(B)}\right)^2 \hat{T}_1 &=& J(J+1)\hat{T}_1, \nonumber\\ 
	\hat{J}_z^{(B)}\hat{T}_1 &=& M \hat{T}_1, \nonumber\\
	\hat{T}_1 \hat{J}_z^{(B)} &=& M'\hat{T}_1,
\end{eqnarray}
The same argument can be repeated for any other grouping of indices into sets $A$ and $B$. As a consequence, the proposed state Eq.~(\ref{eqn:form}) satisfies all three requirements contained in Eq.~(\ref{eqn:props}) and is furthermore fully permutation invariant. Finally, we note that this form of the PI states can be visualized from the number of single-box and double-box columns in the Young tableaux for a given irreducible representation of SU(2)~\cite{Georgi1999,Pfeifer2003}.

\section{Analytical results for steady-state population distribution}
\label{app:pop_an}

In this section, we derive the steady state population distribution in specific parameter regimes where an analytic solution is tractable.

\subsubsection{Steady-state population for $\gamma=0$, $w\rightarrow 0$.}

Quantitatively, the $w\rightarrow0$ regime is characterized by the condition that $Nw/\Gamma_c \ll 1$. In this limit, any state $\ket{J,M}$ with $M>-J$ decays rapidly at rate $\Gamma_c$ or faster and is therefore negligibly populated. Therefore, the population dynamics can be studied by considering only the states of the form $\ket{J,-J}$. 

We label the population in a $\ket{J,-J}$ state by $P_J$. Probability conservation implies that $\sum_{J=0}^{N/2} P_J = 1$. The repump causes transitions from $(J,-J)\rightarrow(J-1,-J+1)$ and $(J,-J)\rightarrow(J+1,-J+1)$ with respective rates $R_J^-$ and $R_J^+$. However, any state with $M\neq -J$ rapidly decays to $M=-J$, therefore the latter transition is effectively $\ket{J,-J}\rightarrow\ket{J+1,-J-1}$ at rate $R_J^+$ in the limit that $Nw/\Gamma_c \ll 1$. Therefore, the rate equation for $P_J$ can be expressed as 
\begin{equation} \label{eqn:dpdt}
	\frac{d P_J}{dt} = -(R_J^- + R_J^+) P_J + R_{J-1}^+ P_{J-1} + R_{J+1}^- P_{J+1}.
\end{equation}
For the highest $J$ value ($J=N/2$), there is no coupling possible to a $J+1$ state, so the $R_J^+$ and $R_{J+1}^-$ terms do not enter the expression for $\frac{d P_J}{dt}$ in Eq.~\eqref{eqn:dpdt} with $J=N/2$. Therefore, the steady-state population for $J=N/2$ is given by 
\begin{equation}
	P_{N/2} = \frac{R_{N/2-1}^+}{R_{N/2}^-} P_{N/2-1}.
\end{equation}
By using this expression as a starting point, we can iteratively relate $P_{J}$ to $P_{J-1}$ and obtain the simple relation 
\begin{equation}
	P_J = \frac{R_{J-1}^+}{R_J^-}P_{J-1}.
	\label{eqn:pj_recursion}
\end{equation}
Therefore, all the $P_J$ values can be recursively expressed in terms of the single population $P_0$ as 
\begin{equation}
	P_J = \left(\Pi_{k=1}^J \frac{R_{k-1}^+}{R_k^-}\right)P_0.
	\label{eqn:pop_recur}
\end{equation}
From the probability conservation, we can then solve for $P_0$ as 
\begin{equation}
	P_0 = \frac{1}{1 + \sum_{J=1}^{N/2} \left(\Pi_{k=1}^J \frac{R_{k-1}^+}{R_k^-}\right) }.
\end{equation}
Since we have the constraint $M=-J$, the rates $R_J^\pm$ depend only on the $J$ values and can be read off from Table~\ref{tab:rates}. We find that 
\begin{equation}
	\frac{R_{J-1}^+}{R_J^-} = \left(\frac{N-2J+2}{N+2J+2}\right)\frac{2J+1}{J(2J-1)^2}.
\end{equation}	
We consider the case when $N\gg 1$. For $J\sim\mathcal{O}(N/2)$, the ratio $\sim 1/N^3$, therefore there is negligible population in the large $J$ states. For small $J$, we take the $N\rightarrow\infty$ limit and obtain   
\begin{equation}\label{eqn:J}
	\frac{R_{J-1}^+}{R_J^-} \approx \frac{2J+1}{J(2J-1)^2}.
\end{equation}
\begin{table}[b]
\begin{tabular}{cl}
	$J$ & $\dfrac{R_{J-1}^+}{R_J^-}$ \\ \hline
	$1$	&	$3$ \\[0.75em]
	$2$	&	$\dfrac{5}{18}\approx0.278$ \\[0.75em] 
	$3$ &	$\dfrac{7}{75}\approx0.093$ \\[0.75em]
	$4$	&	$\dfrac{9}{196}\approx0.046$	\\[0.75em] 
\end{tabular}
\caption{The ratio $R_{J-1}^+/R_{J}^-$ evaluated for the first few $J$ values, cf. Eq~\eqref{eqn:J}.}
\label{tab:ratio_vals}
\end{table}
In Table~\ref{tab:ratio_vals}, we evaluate this ratio for the first few $J$ values. Clearly, this ratio decreases rapidly as $J$ increases, therefore, we only consider the first three $J$ values ($J=0,1,2$) to estimate the steady-state populations in the $N\rightarrow\infty$ limit. Therefore, $P_0,P_1,P_2$ can be readily evaluated to give 
\begin{eqnarray}
	P_0 &\approx& \frac{1}{1+3(1+\frac{5}{18})} = \frac{6}{29} \approx 0.207 \nonumber\\ 
	P_1 &=& 3 P_0 \approx 0.621\nonumber\\
	P_2 &=& \frac{5}{18}P_1 \approx 0.172.
\end{eqnarray}
The neglected population $P_3 = 7P_2/75 \approx 0.016 \ll 1$, therefore these steady-state population estimates are reasonably accurate.

\subsubsection{Steady-state population for $0<w<\gamma$}

Numerically, we find that only the $M=-J$ states are occupied even when $\gamma>0$, provided that $0<w<\gamma$. Therefore, we can restrict our analysis to the $M=-J$ states as in the previous section.

In addition to transitions caused by the repump, the individual atom decay causes transitions from $(J,-J)\rightarrow(J+1,-J-1)$. The corresponding rate is given by 
\begin{eqnarray}
    G_J^+ = \gamma\left[\frac{N-2J}{2} \right].
\end{eqnarray}
The rate equations from the previous section are now applicable with the modification 
\begin{eqnarray}
    R_J^+ \rightarrow R_J^+ + G_J^+.
\end{eqnarray}
The recursive expression, Eq.~(\ref{eqn:pop_recur}), for $P_J$ is now valid with the modified ratio 
\begin{eqnarray}
    \frac{R_{J-1}^+}{R_J} = \left(\frac{N-2J-2}{N+2J+2}\right)\left[\frac{\frac{1}{J(2J-1)} + \frac{\gamma}{w}}{\frac{2J-1}{2J+1}}\right].
\end{eqnarray}
For $\gamma \gg w$, we expect the population to be concentrated in $J \sim N/2$. We therefore assume that for a finite $\gamma/w>1$, the mean value of $J\sim\mathcal{O}(N)$. We consider $N\gg 1$, parametrize $J=Nj$ and estimate  $R_{J-1}^+/R_J$ assuming $N,J\gg 1$. The result is 
\begin{eqnarray}
    \frac{R_{J-1}^+}{R_J} \approx \frac{\gamma}{w}\left(\frac{1-2j}{1+2j}\right) \equiv r(j).
\end{eqnarray}
We have defined a function $r(j)$ anticipating a continuum approximation of the discrete-valued $j$ in the large $N$ limit. 

The value $j_0$ satisfying $r(j_0)=1$ corresponds to the peak of the population distribution since for $j>j_0$ the ratio $r(j)<1$ and the population will start decreasing in value according to Eq.~(\ref{eqn:pop_recur}). The value of $j_0$ is found to be 
\begin{eqnarray}
j_0 = \frac{1}{2}\left[\frac{\gamma/w-1}{\gamma/w+1} \right].    
\end{eqnarray}

From Eq.~(\ref{eqn:pop_recur}), the population $P_J$ is proportional to 
\begin{eqnarray}
    P_J \propto \exp\left[\sum_{k=1}^J \log\left(\frac{R_{k-1}^+}{R_k^-}\right) \right].
\end{eqnarray}
Taking the continuum limit and changing variables to $j=J/N$, we get 
\begin{eqnarray}
    P(j) \propto \exp\left[N\int_0^j dx \log(r(x)) \right].
\end{eqnarray}
We consider the argument of the exponential  
\begin{eqnarray}
    f(j) = N\int_0^j dx \log (r(x))
\end{eqnarray}
and expand it around its maximum value which occurs at $j=j_0$. The result is 
\begin{eqnarray}
    f(j) \approx Nf(j_0) + \frac{N f''(j_0)}{2} (j-j_0)^2,
\end{eqnarray}
where the second derivative $f''(j) = r'(j)/r(j)$ at $j=j_0$ evaluates to
\begin{eqnarray}
f''(j_0) = -\frac{(1+\gamma/w)^2}{\gamma/w}.    
\end{eqnarray}
Therefore, the probability distribution is a Gaussian and is given by 
\begin{eqnarray}
p(j) \propto \exp\left[-\frac{N\abs{f''(j_0)}}{2}(j-j_0)^2 \right].
\end{eqnarray}
For $1<\gamma/w<\infty$, where this solution is valid, the center of the distribution $j_0$ monotonically decreases from $1/2$ to $0$ as $\gamma/w$ decreases, while the variance in $j$ steadily increases from $0$ to $1/(4N)$. 

\bibliography{references.bib}

@article{dicke1954PR,
  title = {Coherence in Spontaneous Radiation Processes},
  author = {Dicke, R. H.},
  journal = {Phys. Rev.},
  volume = {93},
  issue = {1},
  pages = {99--110},
  numpages = {0},
  year = {1954},
  month = {Jan},
  publisher = {American Physical Society},
  doi = {10.1103/PhysRev.93.99},
  url = {https://link.aps.org/doi/10.1103/PhysRev.93.99}
}

@article{meiser2009PRL,
  title = {Prospects for a Millihertz-Linewidth Laser},
  author = {Meiser, D. and Ye, Jun and Carlson, D. R. and Holland, M. J.},
  journal = {Phys. Rev. Lett.},
  volume = {102},
  issue = {16},
  pages = {163601},
  numpages = {4},
  year = {2009},
  month = {Apr},
  publisher = {American Physical Society},
  doi = {10.1103/PhysRevLett.102.163601},
  url = {https://link.aps.org/doi/10.1103/PhysRevLett.102.163601}
}

@article{meiser2010PRASteadyState,
  title = {Steady-state superradiance with alkaline-earth-metal atoms},
  author = {Meiser, D. and Holland, M. J.},
  journal = {Phys. Rev. A},
  volume = {81},
  issue = {3},
  pages = {033847},
  numpages = {4},
  year = {2010},
  month = {Mar},
  publisher = {American Physical Society},
  doi = {10.1103/PhysRevA.81.033847},
  url = {https://link.aps.org/doi/10.1103/PhysRevA.81.033847}
}

@article{urizar2013PRA,
  title = {Macroscopic singlet states for gradient magnetometry},
  author = {Urizar-Lanz, Iñigo and Hyllus, Philipp and Egusquiza, Iñigo Luis and Mitchell, Morgan W. and T\'oth, G\'eza},
  journal = {Phys. Rev. A},
  volume = {88},
  issue = {1},
  pages = {013626},
  numpages = {13},
  year = {2013},
  month = {Jul},
  publisher = {American Physical Society},
  doi = {10.1103/PhysRevA.88.013626},
  url = {https://link.aps.org/doi/10.1103/PhysRevA.88.013626}
}

@article{zhang2018NJP,
	doi = {10.1088/1367-2630/aaec36},
	url = {https://doi.org/10.1088%2F1367-2630%2Faaec36},
	year = 2018,
	month = {nov},
	publisher = {{IOP} Publishing},
	volume = {20},
	number = {11},
	pages = {112001},
	author = {Yuan Zhang and Yu-Xiang Zhang and Klaus M{\o}lmer},
	title = {Monte-Carlo simulations of superradiant lasing},
	journal = {New Journal of Physics}
}

@article{ostermann2013PRL,
  title = {Protected State Enhanced Quantum Metrology with Interacting Two-Level Ensembles},
  author = {Ostermann, Laurin and Ritsch, Helmut and Genes, Claudiu},
  journal = {Phys. Rev. Lett.},
  volume = {111},
  issue = {12},
  pages = {123601},
  numpages = {5},
  year = {2013},
  month = {Sep},
  publisher = {American Physical Society},
  doi = {10.1103/PhysRevLett.111.123601},
  url = {https://link.aps.org/doi/10.1103/PhysRevLett.111.123601}
}

@article{temnov2009OptExp,
author = {Vasily V. Temnov and Ulrike Woggon},
journal = {Opt. Express},
number = {7},
pages = {5774--5782},
publisher = {OSA},
title = {Photon statistics in the cooperative spontaneous emission},
volume = {17},
month = {Mar},
year = {2009},
url = {http://www.opticsexpress.org/abstract.cfm?URI=oe-17-7-5774},
doi = {10.1364/OE.17.005774},
}

@book{mandel1995Optical, place={Cambridge}, title={Optical Coherence and Quantum Optics}, DOI={10.1017/CBO9781139644105}, publisher={Cambridge University Press}, author={Mandel, Leonard and Wolf, Emil}, year={1995}}

@article{shankar2021PRL,
  title = {Subradiant-to-Subradiant Phase Transition in the Bad Cavity Laser},
  author = {Shankar, Athreya and Reilly, Jarrod T. and J\"ager, Simon B. and Holland, Murray J.},
  journal = {Phys. Rev. Lett.},
  volume = {127},
  issue = {7},
  pages = {073603},
  numpages = {6},
  year = {2021},
  month = {Aug},
  publisher = {American Physical Society},
  doi = {10.1103/PhysRevLett.127.073603},
  url = {https://link.aps.org/doi/10.1103/PhysRevLett.127.073603}
}

@Article{cocconi2020Entropy,
AUTHOR = {Cocconi, Luca and Garcia-Millan, Rosalba and Zhen, Zigan and Buturca, Bianca and Pruessner, Gunnar},
TITLE = {Entropy Production in Exactly Solvable Systems},
JOURNAL = {Entropy},
VOLUME = {22},
YEAR = {2020},
NUMBER = {11},
ARTICLE-NUMBER = {1252},
URL = {https://www.mdpi.com/1099-4300/22/11/1252},
ISSN = {1099-4300},
DOI = {10.3390/e22111252}
}

@article{chase2008PRA,
  title = {Collective processes of an ensemble of spin-$1/2$ particles},
  author = {Chase, Bradley A. and Geremia, J. M.},
  journal = {Phys. Rev. A},
  volume = {78},
  issue = {5},
  pages = {052101},
  numpages = {14},
  year = {2008},
  month = {Nov},
  publisher = {American Physical Society},
  doi = {10.1103/PhysRevA.78.052101},
  url = {https://link.aps.org/doi/10.1103/PhysRevA.78.052101}
}

@article{edgar2015PRL,
  title = {Decision Making in the Arrow of Time},
  author = {Rold\'an, \'Edgar and Neri, Izaak and D\"orpinghaus, Meik and Meyr, Heinrich and J\"ulicher, Frank},
  journal = {Phys. Rev. Lett.},
  volume = {115},
  issue = {25},
  pages = {250602},
  numpages = {5},
  year = {2015},
  month = {Dec},
  publisher = {American Physical Society},
  doi = {10.1103/PhysRevLett.115.250602},
  url = {https://link.aps.org/doi/10.1103/PhysRevLett.115.250602}
}

@article{seifert2005PRL,
  title = {Entropy Production along a Stochastic Trajectory and an Integral Fluctuation Theorem},
  author = {Seifert, Udo},
  journal = {Phys. Rev. Lett.},
  volume = {95},
  issue = {4},
  pages = {040602},
  numpages = {4},
  year = {2005},
  month = {Jul},
  publisher = {American Physical Society},
  doi = {10.1103/PhysRevLett.95.040602},
  url = {https://link.aps.org/doi/10.1103/PhysRevLett.95.040602}
}

@article{landi2021RMP,
  title = {Irreversible entropy production: From classical to quantum},
  author = {Landi, Gabriel T. and Paternostro, Mauro},
  journal = {Rev. Mod. Phys.},
  volume = {93},
  issue = {3},
  pages = {035008},
  numpages = {58},
  year = {2021},
  month = {Sep},
  publisher = {American Physical Society},
  doi = {10.1103/RevModPhys.93.035008},
  url = {https://link.aps.org/doi/10.1103/RevModPhys.93.035008}
}

@article{santos2017PRL,
  title = {Wigner Entropy Production Rate},
  author = {Santos, Jader P. and Landi, Gabriel T. and Paternostro, Mauro},
  journal = {Phys. Rev. Lett.},
  volume = {118},
  issue = {22},
  pages = {220601},
  numpages = {5},
  year = {2017},
  month = {Jun},
  publisher = {American Physical Society},
  doi = {10.1103/PhysRevLett.118.220601},
  url = {https://link.aps.org/doi/10.1103/PhysRevLett.118.220601}
}

@article{santos2018PRA,
  title = {Spin-phase-space-entropy production},
  author = {Santos, Jader P. and C\'eleri, Lucas C. and Brito, Frederico and Landi, Gabriel T. and Paternostro, Mauro},
  journal = {Phys. Rev. A},
  volume = {97},
  issue = {5},
  pages = {052123},
  numpages = {10},
  year = {2018},
  month = {May},
  publisher = {American Physical Society},
  doi = {10.1103/PhysRevA.97.052123},
  url = {https://link.aps.org/doi/10.1103/PhysRevA.97.052123}
}

@article{goes2020PRR,
  title = {Quantum features of entropy production in driven-dissipative transitions},
  author = {Goes, Bruno O. and Fiore, Carlos E. and Landi, Gabriel T.},
  journal = {Phys. Rev. Research},
  volume = {2},
  issue = {1},
  pages = {013136},
  numpages = {9},
  year = {2020},
  month = {Feb},
  publisher = {American Physical Society},
  doi = {10.1103/PhysRevResearch.2.013136},
  url = {https://link.aps.org/doi/10.1103/PhysRevResearch.2.013136}
}

@article{PhysRevA87,
  title = {Simulating open quantum systems by applying SU(4) to quantum master equations},
  author = {Xu, Minghui and Tieri, D. A. and Holland, M. J.},
  journal = {Phys. Rev. A},
  volume = {87},
  issue = {6},
  pages = {062101},
  numpages = {7},
  year = {2013},
  month = {Jun},
  publisher = {American Physical Society},
  doi = {10.1103/PhysRevA.87.062101},
  url = {https://link.aps.org/doi/10.1103/PhysRevA.87.062101}
}

@article{PhysRevA98,
  title = {Open quantum systems with local and collective incoherent processes: Efficient numerical simulations using permutational invariance},
  author = {Shammah, Nathan and Ahmed, Shahnawaz and Lambert, Neill and De Liberato, Simone and Nori, Franco},
  journal = {Phys. Rev. A},
  volume = {98},
  issue = {6},
  pages = {063815},
  numpages = {34},
  year = {2018},
  month = {Dec},
  publisher = {American Physical Society},
  doi = {10.1103/PhysRevA.98.063815},
  url = {https://link.aps.org/doi/10.1103/PhysRevA.98.063815}
}

@article{PhysRevLett114,
  title = {Conditional Ramsey Spectroscopy with Synchronized Atoms},
  author = {Xu, Minghui and Holland, M. J.},
  journal = {Phys. Rev. Lett.},
  volume = {114},
  issue = {10},
  pages = {103601},
  numpages = {5},
  year = {2015},
  month = {Mar},
  publisher = {American Physical Society},
  doi = {10.1103/PhysRevLett.114.103601},
  url = {https://link.aps.org/doi/10.1103/PhysRevLett.114.103601}
}

@article{Gaspard,
    author = {Gaspard, Pierre},
    title = {Time-Reversed Dynamical Entropy and Irreversibility in Markovian Random Processes},
    journal = {Journal of Statistical Physics},
    year = {2004},
    pages = {599},
    doi = {10.1007/s10955-004-3455-1}
}

@article{Seifert_2012,
doi = {10.1088/0034-4885/75/12/126001},
url = {https://dx.doi.org/10.1088/0034-4885/75/12/126001},
year = {2012},
month = {nov},
publisher = {IOP Publishing},
volume = {75},
number = {12},
pages = {126001},
author = {Seifert, Udo},
title = {Stochastic thermodynamics, fluctuation theorems and molecular machines},
journal = {Reports on Progress in Physics}
}

@article{esposito,
  title = {Three faces of the second law. I. Master equation formulation},
  author = {Esposito, Massimiliano and Van den Broeck, Christian},
  journal = {Phys. Rev. E},
  volume = {82},
  issue = {1},
  pages = {011143},
  numpages = {10},
  year = {2010},
  month = {Jul},
  publisher = {American Physical Society},
  doi = {10.1103/PhysRevE.82.011143},
  url = {https://link.aps.org/doi/10.1103/PhysRevE.82.011143}
}

@article{jaeger2022PRL,
  title = {Lindblad Master Equations for Quantum Systems Coupled to Dissipative Bosonic Modes},
  author = {J\"ager, Simon B. and Schmit, Tom and Morigi, Giovanna and Holland, Murray J. and Betzholz, Ralf},
  journal = {Phys. Rev. Lett.},
  volume = {129},
  issue = {6},
  pages = {063601},
  numpages = {7},
  year = {2022},
  month = {Aug},
  publisher = {American Physical Society},
  doi = {10.1103/PhysRevLett.129.063601},
  url = {https://link.aps.org/doi/10.1103/PhysRevLett.129.063601}
}

@phdthesis{xu2016theory,
  title={Theory of steady-state superradiance},
  author={Xu, Minghui},
  year={2016},
  school={University of Colorado at Boulder}
}

@article{baragiola2010PRA,
  title = {Collective uncertainty in partially polarized and partially decohered spin-$\frac{1}{2}$ systems},
  author = {Baragiola, Ben Q. and Chase, Bradley A. and Geremia, JM},
  journal = {Phys. Rev. A},
  volume = {81},
  issue = {3},
  pages = {032104},
  numpages = {12},
  year = {2010},
  month = {Mar},
  publisher = {American Physical Society},
  doi = {10.1103/PhysRevA.81.032104},
  url = {https://link.aps.org/doi/10.1103/PhysRevA.81.032104}
}

@article{reilly2026PRL,
  title = {Fully Collective Superradiant Lasing with Vanishing Sensitivity to Cavity Length Vibrations},
  author = {Reilly, Jarrod T. and J\"ager, Simon B. and Cooper, John and Holland, Murray J.},
  journal = {Phys. Rev. Lett.},
  volume = {136},
  issue = {14},
  pages = {143803},
  numpages = {7},
  year = {2026},
  month = {Apr},
  publisher = {American Physical Society},
  doi = {10.1103/v6jq-m6sk},
  url = {https://link.aps.org/doi/10.1103/v6jq-m6sk}
}

@book{Georgi1999,
  title = {Lie Algebras In Particle Physics: from Isospin To Unified Theories},
  author = {Georgi, H.},
  isbn = {9780738202334},
  lccn = {99064878},
  series = {Frontiers in Physics},
  year = {1999},
  publisher = {Avalon Publishing},
  url = {https://books.google.com/books?id=p58_BAAAQBAJ}
}

@book{Pfeifer2003, 
  author = {Pfeifer, Walter}, 
  title = {The Lie Algebras su(N), An Introduction},
  year = {2003},
  place = {Boston, MA},  
  publisher = {Birkh\"auser},
  url = {https://link.springer.com/book/10.1007/978-3-0348-8097-8}
}

@misc{dataset,
key = {},
  title        = {Dynamical aspects of steady-state subradiance: Detailed balance and its breakdown (Code to generate data)},
  year         = {2026},
  url          = {https://doi.org/10.5281/zenodo.20192908}
}

\end{document}